\documentclass[preprint,12pt]{elsarticle}
\usepackage{amsmath,amssymb}
\usepackage{graphicx}
\usepackage{xcolor}
\usepackage[hidelinks]{hyperref}
\biboptions{sort&compress}
\newcommand{\re}{\operatorname{Re}}
\newcommand{\im}{\operatorname{Im}}
\newcommand{\sech}{\operatorname{sech}}
\newcommand{\dd}{\mathrm{d}}

\journal{Wave Motion}
\begin{document}
\begin{frontmatter}

\title{Hamiltonian theory of motion of bright solitons and breathers of the focusing nonlinear Schr\"odinger equation along a large-scale background}

\author[a,b]{D.V. Shaykin\corref{cor1}}
\ead{shaykin.dv@phystech.edu}
\cortext[cor1]{Corresponding author.}
\affiliation[a]{organization={Russian University of Transport (RUT-MIIT)},
            city={Moscow},
            country={Russia,},postcode={127994}}
 \affiliation[b]{{Moscow Institute of Physics and Technology},
            city={Dolgoprudny},
            country={Russia,},postcode={141700}}

\begin{abstract}
The motion of bright solitons (breathers) of the NLS equation along smooth evolving background waves is studied. Considering the tail of the soliton as a linear wave with a complex frequency and wave number, the law of motion and the conservation law for the complex parameter of the soliton are established. The explicit form of the Hamiltonian and canonical variables of the background wave and soliton characteristics through a single complex angle has been found. It is shown that along the path of the soliton, the sum of the norm of the soliton and its wake left in the background wave is conserved. Numerical simulations of solitons passing through a parabolic hill agree well with the theory. It has been shown that the phase shift acquired by the particle on the slope is carried along its trajectory.
\end{abstract}

\begin{keyword}
nonlinear Schr\"odinger equation \sep breathers \sep soliton dynamics \sep wave packets \sep Hamiltonian mechanics 
\end{keyword}

\end{frontmatter}

\section{Introduction}

The motion of a soliton along a smooth, large-scale background wave is one of the main problems of dispersive hydrodynamics. The interaction of solitons with rarefaction waves and dispersive shock waves, including soliton trapping and ``hydrodynamic tunneling'', has been theoretically studied and experimentally observed~\cite{ElHoefer2016,Maiden2018,Sprenger2018,Ablowitz2023}. In these works, one-parameter solitons of the Korteweg--de Vries equation or the defocusing nonlinear Schr\"odinger equation were studied.

For the bright soliton of the focusing NLS equation
\begin{equation}\label{int1}
  i\psi_t+\tfrac12\psi_{xx}+|\psi|^2\psi=0
\end{equation}
Zakharov and Manakov constructed action-angle variables for equation~\eqref{int1} on a zero background, in which the soliton is described by two canonical pairs: its position and momentum, as well as its phase and norm~\cite{ZakharovManakov1974}. Perturbation theory based on inverse scattering (IST)~\cite{KarpmanMaslov1977,KaupNewell1978,KivsharMalomed1989} or on the variational ansatz~\cite{Anderson1983} provides Hamiltonian equations for a soliton in a weak or slowly varying external potential. In these theories, the inhomogeneity is a given field, not a background wave with its own dynamics, the norm of the soliton is conserved, and the soliton does not exchange norm with the surrounding medium.

A different approach, which needs neither the IST nor an explicit perturbation theory, uses Stokes's remark~\cite{Stokes1905} that the exponentially small tails of a soliton obey the same linearized equation as small-amplitude harmonic waves. Consequently, the soliton velocity follows from the linear dispersion relation $\omega(k)$ continued to imaginary wave numbers, $k\mapsto i\kappa$, where $\kappa$ is the inverse half-width of the soliton.  It has been shown that there is a conservation law for the real \( k \) (for the linear packet~\cite{ShaykinKamchatnov2023}) and its imaginary extension \( ik \) (for the soliton \cite{KamchatnovShaykin2023PRE,Kamchatnov2024TMF,KamchatnovShaykin2024WM,Kamchatnov2024SAPM,Kamchatnov2025PRE}), based on which Hamiltonian theories of one-parameter solitons were constructed. The bright NLS soliton on a nonzero background is not of this type. Its tails oscillate against the background, so the tail wave number $K$ and frequency $\Omega$ are complex, and the soliton has two real parameters. Families of such two-parameter solutions on a uniform background are known exactly: the Kuznetsov--Ma~\cite{Kuznetsov1977,Ma1979}, Akhmediev~\cite{AkhmedievKorneev1986} and Peregrine~\cite{Peregrine1983} breathers, and the general moving breather of Tajiri and Watanabe~\cite{TajiriWatanabe1998}. Their spectral theory is given by the IST with nonzero boundary conditions~\cite{BiondiniKovacic2014}. The kinematic side of Stokes's remark was extended to complex wave numbers by Sazonov~\cite{KozlovSazonov1997,Sazonov2018} (the method of analytic
continuation of dispersion parameters): the envelope of a two-parameter
soliton moves with the velocity $V=\operatorname{Im}\Omega/\operatorname{Im}K$ and its carrier with $\operatorname{Re}\Omega/\operatorname{Re}K$. These results, however, refer to a uniform medium in which the soliton parameters are constant.

Soliton motion through a nonuniform focusing background was studied by Biondini, Li and Mantzavinos~\cite{BiondiniLiMantzavinos2018}, who found, by the IST and direct numerical simulations, that a soliton can be transmitted through or trapped by the structures produced by modulational instability, that its velocity changes inside them, and that it can leave a wake behind. This analysis is carried out for specific initial data and gives no quantitative rule for how much norm the soliton leaves in the background.

In this work, we develop a Hamiltonian theory of a two-parameter soliton moving along a non-uniform evolving wave. Since the focusing NLS is modulationally unstable, we assume that during the motion of the soliton it does not encounter the developing instability, or that the soliton crosses the nonuniform region in a time much shorter than the growth time of the instability.

\section{Equations of motion}\label{sec:kin}
\subsection{Background and linear waves}
A large-scale background $\psi_b=\sqrt{\rho}\,e^{i\vartheta}$, $ u=\vartheta_x$, obeys the dispersionless equations
\begin{equation}\label{eq1}
\rho_t+(\rho u)_x=0,\qquad  u_t+ u u_x-\rho_x=0 ,
\end{equation}
which are diagonalized by the complex Riemann invariant
\begin{equation}\label{eq2}
r=\frac{ u}2+i\sqrt{\rho},\qquad r_t+v\,r_x=0,\qquad v=\tfrac12(3r+r^*)
\end{equation}
and the corresponding conjugate expression. The complex characteristic velocity reflects the modulational instability of the focusing background; the dispersionless description is valid on time scales shorter than $\sim 1/\rho$, the inverse maximal growth rate (see the dispersion expression below). The inverse substitution leads to the physical variables $u=r+r^*$ and $2i\sqrt{\rho} = r-r^*$.

Linear waves $\propto e^{i(kx-\omega t)}$ on a uniform background obey
\begin{equation}\label{eq3}
\omega=uk\pm k \sqrt{k^2/4-\rho}.
\end{equation}
A wave packet on a large-scale background moves according to Hamiltonian equations $\dot x=\partial\omega/\partial k$, $\dot k=-\partial\omega/\partial x$. These equations admit
the integral~\cite{ShaykinKamchatnov2023,Kamchatnov2025PRE}
\begin{equation}\label{eq4}
k^2=4(q-r)(q-r^*),\qquad \frac\omega k=\frac{ u}2+q,
\end{equation}
where $q$ is constant along the motion. The second relation is derived  from eq. (\ref{eq3}), the first shows that the wave number depends only on the local background.

\subsection{Motion of solitons}
The tail of the bright soliton
\begin{equation}\label{eq5}
    \psi=\kappa\sech\big[\kappa(x-Vt)\big]e^{iVx+\frac i2(\kappa^2-V^2)t}
\end{equation}
at $x\to+\infty$ is $\propto e^{i(Kx-\Omega t)}$ with $K=V+i\kappa$, $\Omega=K^2/2$, i.e., a linear
wave with complex wave number and frequency. The modulus of the tail is $\exp[-\im K(x-t\im\Omega/\im K)]$, so the soliton moves with the
velocity~\cite{KozlovSazonov1997,Sazonov2018}
\begin{equation}\label{eq6}
V=\frac{\im\Omega}{\im K}.
\end{equation}
The tails of breathers on a finite background are linear waves (\ref{eq3}) with complex $K$ and $\Omega$ as well. Applying (\ref{eq4}) to them, we obtain
\begin{equation}\label{eq7}
K^2=4(q-r)(q-r^*),\qquad \Omega=K\Bigl(\frac{u}2+q\Bigr),
\end{equation}
with a complex constant $q$ along the motion. We take the branch of the root with $\im K>0$, which corresponds to all standard breathers. Equations (\ref{eq6}) and (\ref{eq7}) then give the equations of motion
\begin{equation}\label{eq8}
\dot q=0,\qquad \frac{\dd x}{\dd t}=V=\frac{u}2+\frac{\im(Kq)}{\im K},
\end{equation}
where $K=2\sqrt{(q-r)(q-r^*)}$ and $u$, $r$, $r^*$ are taken at the soliton position. On a zero background $K=2q-u$, so $2q=V+i\kappa$ for the bright soliton. On a uniform background $u=0$, $\rho=a^2$, the choice $q=i\eta$ with $\eta>a$ gives $V=0$, i.e., the Kuznetsov--Ma breather, and a general complex $q$ gives the moving Tajiri--Watanabe breather.

In the soliton frame $\xi=x-Vt$ the tail factor is $e^{iK\xi}e^{-i(\Omega-VK)t}$, and $\Omega-VK$ is real by (\ref{eq6}). Hence the tail rotates in phase relative to the background, and the density in the tail oscillates as $e^{-\im K\xi}\cos(\re K\xi-\omega_bt)$ with the breathing frequency
\begin{equation}\label{eq9}
\omega_b=\re\Omega-V\re K = \Omega-VK.
\end{equation}
This is the beat frequency between the carriers of the soliton and of the background in the soliton frame. For the Kuznetsov--Ma breather (\ref{eq9}) gives $\omega_b=-2\eta\sqrt{\eta^2-a^2}$, i.e.\ the known frequency, and on a zero background $\omega_b=-\frac12(V^2+\kappa^2)$.

\section{Hamiltonian structure}\label{sec:ham}
\subsection{Canonical variables}
As seen from the analysis above, a two-parameter soliton has two degrees of freedom: its position and the phase of the breathing. Following the theory for a zero background~\cite{ZakharovManakov1974}, the pairs of conjugate variables are coordinate and momentum \((x,P)\) and phase and norm \((\theta,N)\). We look for a Hamiltonian that gives the equations of motion:
\begin{equation}\label{ham1}
\dot q=0,\qquad \dot x=V,\qquad \dot\theta=\omega_b.
\end{equation}
Although in classical mechanics it is customary to consider \( H = H(x,P,\theta,N) \), it will be convenient for us to seek it in the form \( H = H(q,q^*,r,r^*), P = P(q,q^*,r,r^*), N = N(q,q^*,r,r^*) \), where the two real components of $q$ are replaced by the pair $q,q^*$, and also introduced the redefinitions \( r = r_+, r^* = r_- \) and \( q = q_+, q^* = q_- \) so that all further formulas can be written in a very concise form. Note that all quantities $H,P,N,\theta$ are real.

If we consider the soliton as a particle in the background, its Hamiltonian cannot depend on the phase $\theta$, otherwise the energy would oscillate even for constant parameters of the medium. Therefore $\theta$ is a cyclic variable, and the conjugate variable $N$ is conserved along the motion, $\dot N = 0$. Considering that $\dot N = \sum_{\pm} N_{r_\pm}\dot r_\pm + N_{q_\pm}\dot q_\pm$, and that $q_\pm$ is conserved along the motion, we obtain that $N$ does not depend on the background evolution, i.e., $N = N(q_\pm)$. As in the case of the dark soliton (see ~\ref{A:dark}), from the Hamiltonian equations we will try to find $H_{q_\pm}$ and $H_{r_\pm}$, so that by taking cross derivatives, we can obtain an expression for the momentum equation.

At a fixed background, the equations $\partial H/\partial P|_N=V$ and $\partial H/\partial N|_P=\omega_b$ mean $\dd H=V\dd P+\omega_b\dd N$, which gives
\begin{equation}\label{ham2}
H_{q_\pm}=VP_{q_\pm}+\omega_bN_{q_\pm}.
\end{equation}
For the Hamiltonian equation on momentum, all calculations are done easily if we take into account that when $N,P =\mathrm{const}$, from the equality of their differentials to zero follows $0=\sum_{\pm} \bigg( P_{r_\pm}dr_\pm+P_{q_\pm}dq_\pm\bigg)$, $\sum_\pm N_{q_\pm}dq_\pm=0$, then
\begin{equation}\label{ham3}
H_x\Big|_{P,N} = \sum_{\pm}\Big( H_{q_\pm}q_x+H_{r_\pm}\partial_x r_{\pm}  \Big)\Big|_{P,N} = \sum_{\pm}\Big( -VP_{r_\pm} +H_{r_\pm}\Big)\partial_x  r_{\pm}.
\end{equation}
So, such a calculation, taking into account \(\dot P = \sum_{\pm}P_{r_\pm}\big( V-v_\pm\big)\partial_x r_\pm\), reduces the Hamiltonian equation \(\dot P =-\partial H/ \partial x\) to the following relation: 
\begin{equation}\label{ham4}
H_{r_\pm}=v_\pm P_{r_\pm},
\end{equation}
 having a form analogous to a dark soliton (see ~\ref{A:dark}), where \(\pm\) now denotes complex conjugation.

Now let us take the mixed cross derivatives, that is, differentiate formula \eqref{ham2} with respect to $r_\pm$ and formula \eqref{ham4} with respect to $q_\pm$, then we will obtain the following two relations:
\begin{equation}\label{ham5}
    (v_\pm-V)\,\frac{\partial P_{q_\pm}}{\partial r_\pm}=V_{r_\pm}\,P_{q_\pm}+\omega_{b_{r_\pm}}\,N_{q_\pm}, 
\end{equation}
where $N =N(q_\pm)$, which means we have an inhomogeneous first-order equation in $r_\pm$ for the function $P_{q_\pm}$. The solution to the equation can be found as the solution to the homogeneous equation with variation of constants, which after laborious calculations leads to the solution
\begin{equation}\label{ham6}
    P_q=N_qK+c\im K, \qquad P_{q^*}=N_{q^*}K^*+c^*\im K,
\end{equation}
bringing us back to the traditional notation of conjugate quantities. The equality $\partial_{q^*}P_q=\partial_qP_{q^*}$ gives
\begin{equation}\label{ham7}
    2i\bigl(2iN_{qq^*}+c_{q^*}-c^*_q\bigr)\im K=c\,K^*_{q^*}+c^*K_q,
\end{equation}
where $K=2\sqrt{(q-u/2)^2+\rho}$. As a function of $\rho$ (continued analytically), $K_q$ diverges where $K\to 0$, while all other terms remain finite, hence $c=0$. Then $N_{qq^*}=0$, that is $N_q = \mu(q)$. Since \( N(q,q^*) \) is a norm that does not depend on the background, it is sufficient to compare it with the norm \( 2\kappa \) of the bright soliton \eqref{eq5}, from which it immediately follows that \( \mu = -2i \) and 
\begin{equation}
    N = 4\im q.
\end{equation}
Therefore, \( P_q = -2iK, H_q = -2i(VK+\omega_b)=-2i\Omega \) and the canonical momentum and Hamiltonian are expressed as
\begin{equation}\label{ham8}
    \begin{split}
&P=p+p^*=2\re p,\qquad p=-2i\int_r^qK\,\dd q',\\
&H=h+h^*=2\re h,\qquad h=-2i\int_r^q\Omega\,\dd q' .
\end{split}
\end{equation}

\subsection{Explicit expressions}

The integrals are evaluated with the substitution
\begin{equation}\label{ham9}
q=\tfrac12(r+r^*)+\tfrac12(r-r^*)\cos\Phi,\qquad K=2\sqrt{\rho}\,\sin\Phi,
\end{equation}
which removes the square root in $K$. With the complex angle $\Phi$, we obtain formulas that coincide with the formulas for the dark soliton \eqref{dark9}. If we use the explicit expression $\Phi = \alpha + i\beta$, we obtain the real desired functions
\begin{equation}\label{ham10}
\begin{split}
N&=4\sqrt{\rho}\cos\alpha\cosh\beta,\qquad
P=2\rho\,(\sin2\alpha\cosh2\beta-2\alpha),\\
V&= u+\sqrt{\rho}\,\frac{\sin\alpha\cosh2\beta}{\sinh\beta},\qquad
\omega_b=-\rho\Bigl(\sinh2\beta+\frac{2\sin^2\alpha}{\tanh\beta}\Bigr),\\
H=& uP+8\rho^{3/2}\cos\alpha \sinh\beta\big(\sin^2\alpha\cosh^2\beta
-\tfrac13\cos^2\alpha\sinh^2\beta).
\end{split}
\end{equation}
Despite the cumbersome expressions, they provide a clear form for the soliton parameters. Thus, for $|\alpha|<\pi/2$ and $\beta>0$, the expression $K = 2\sqrt{\rho}\sin{\Phi}$ conformally maps this strip to the region $\im K>0$, just as we would want for the soliton. From the formulas \eqref{ham9}, by equating the imaginary and real parts, it follows that for a given background $r$ the angles $\alpha,\beta$ are found from
\begin{align}\label{ham11}
    \cosh{\beta} = \frac{|q-r|+|q-r^*|}{2\im r},\quad \cos{\alpha} = \frac{|q-r^*|-|q-r|}{2\im r},\quad\\ \mathrm{sign}(\alpha) = \mathrm{sign}\re(q-u/2).
\end{align}
As \( r \to 0 \), it can be verified that the entire expression transitions to the canonical expressions \( N=2\kappa, P=2\kappa V, H = \frac{P^2}{2N} - \frac{N^3}{24} \).

Direct calculations using the exact Tajiri--Watanabe breather in our notation (see ~\ref{app:breather}) show that the expressions found above coincide with the renormalized expressions
\begin{align}\label{ham12}
   E=&\int\bigl[\tfrac12|\psi_x|^2-\tfrac12(|\psi|^2-\rho)^2\bigr]\dd x,\\ 
   P=&\int\im(\psi^*\psi_x)\dd x-\rho\,\Delta\vartheta
\end{align}
just as it was for other solitons \cite{tsuzuki1971nonlinear,shevchenko1988quasi,konotop2004landau}. For example, the contribution from the background should be removed from the Hamiltonian, and the contribution from the counterflow should be removed from the momentum, which through direct calculations leads to the physical interpretation $4\alpha=\Delta\vartheta$, the jump of the background phase across the soliton (i.e., $\alpha$ is related to the counterflow). This agrees with the so-called theta condition of the IST with nonzero boundary conditions~\cite{BiondiniKovacic2014}. More important is the difference between the norm of the soliton
\begin{equation}\label{ham13}
    N_s=\int(|\psi|^2-\rho)\,\dd x=2\im K=4\sqrt{\rho}\cos\alpha\sinh\beta
\end{equation}
and the conservation of \(N = 4\im q\) along the path on an inhomogeneous background. This indicates that the soliton exchanges norm with the background wave. The two coincide only on a zero background, where $N_{s}=N=2\kappa$. The conserved quantity conjugate
to the breathing phase is the action $N$, a ``renormalized norm'', and not $N_{s}$; the latter depends on the background through $K(q;r,r^*)$ and therefore changes along the soliton path. As a direct consequence: a soliton that moves from a zero background (norm $2\kappa_0$) to a region where its wave number is $K$ carries the norm $2\im K$, and the difference
\begin{equation}\label{ham14}
\Delta N=2\kappa_0-2\im K
\end{equation}
is left in the background behind it.

\section{An example of motion}

Let us consider an example of a soliton moving along an evolving parabolic hump with the initial condition
\begin{equation}\label{ex1}
 \rho_0(x)=
  \begin{cases}
    a^2(1-x^2/l^2), & \text{   \ $|x|\le l$}, \\
     0,       & \text{   \ $|x|>l$},
  \end{cases}
\end{equation}
called Talanov's hump, the evolution of which in the non-dispersive approximation is described in works \cite{talanov1965self,kamchatnov2021propagation}. The solution of system \eqref{eq1}  with the initial condition \eqref{ex1} has a parametric form \begin{equation}\label{ex2}
\begin{split}
  &\rho_b(x,t)=\frac{a^2}{f(t)}\left(1-\frac{x^2}{l^2f^2(t)}\right),\\
  &u_b(x,t)=\frac{f'(t)}{f(t)}x
  =-\frac{2a}{l}\frac{\sqrt{1-f}}{f^{3/2}}x,
  \end{split}
\end{equation}
where $f(t)$ can be found from 
\begin{equation}\label{ex3}
  (2a/l)t=\sqrt{f(1-f)}+\arccos\sqrt{f}.
\end{equation} 

\begin{figure}[t]
\begin{center}
	\includegraphics[scale = 0.8]{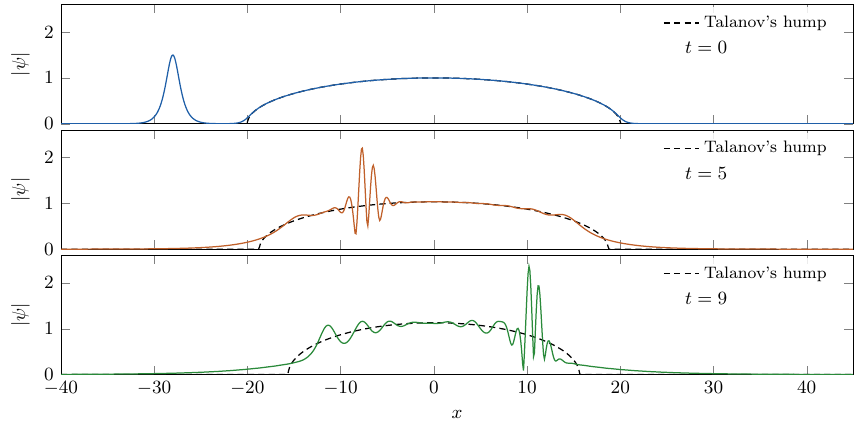}
\caption{The full field $|\psi|$ (solid) and the Talanov dispersionless hump $l=20,a=1$ \eqref{ex2} (dashed) for the soliton $x_0=-28,V_0=4,\kappa_0=1.5$.}
\label{fig1}
\end{center}
\end{figure}

\begin{figure}[t]
\begin{center}
	\includegraphics[scale = 0.8]{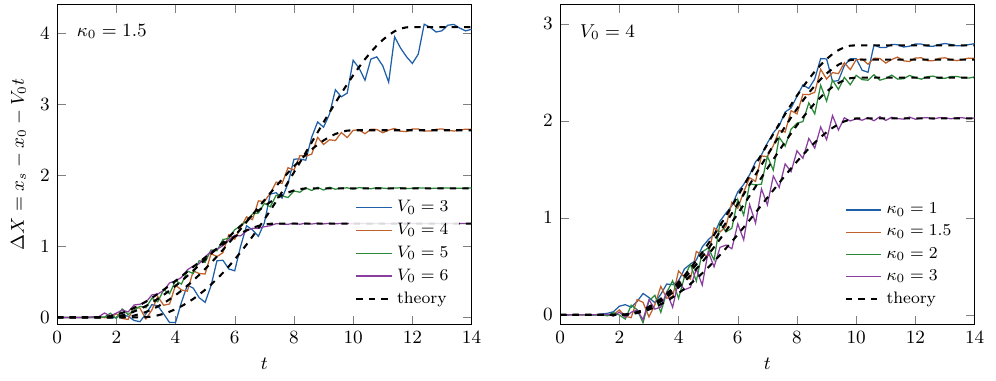}
\caption{Soliton displacement by the Talanov hump \eqref{ex1} with $a=1,l=20$. Left: different velocities at $\kappa_0=1.5$ (the curve for $V_0=3$ deviates at $t\approx10$--$12$, when the hump is already destroyed by the modulation instability, but the final displacement differs from the theory by $0.7\%$); Right: different amplitudes at $V_0=4$.
Solid: numerics; dashed: theory, $\dot x=\mathrm{Im}\,\Omega/\mathrm{Im}\,K$ with $q=(V_0+i\kappa_0)/2=\mathrm{const}$ and the background
from the exact Talanov solution \eqref{ex2}. All solitons start at $x_0=-28$.}
\label{fig2}
\end{center}
\end{figure}

\begin{figure}[t]
\begin{center}
	\includegraphics[scale = 0.8]{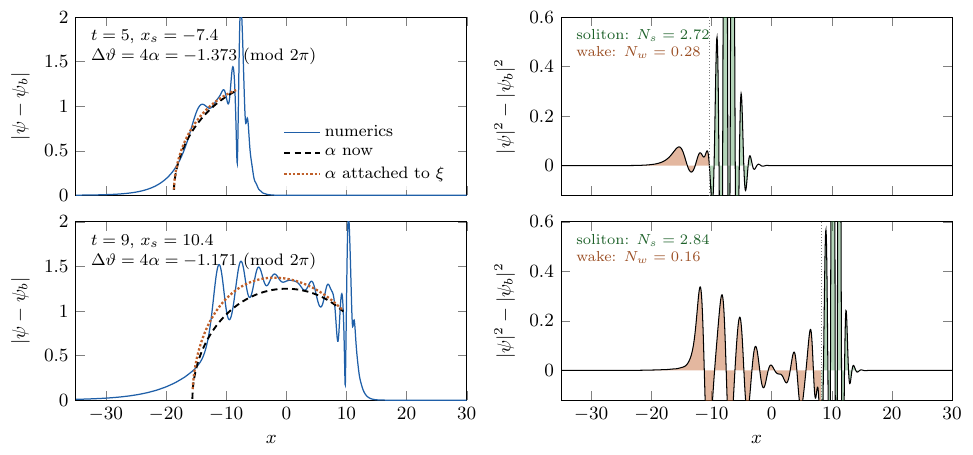}
\caption{A soliton with $V_0=4$, $\kappa_0=1.5$, $x_0=-28$ depicted in Figure \ref{fig1} and its wake in the Talanov hump.
Left: Black dashed: $2\sqrt{\rho(x,t)}\,|\sin2\alpha(t)|$, $\rho$ from the Talanov solution, $\alpha(t) =\alpha(5),\alpha(9)$. Orange dotted: the same with the phase jump $\Delta\vartheta=4\alpha(\tau)$ described in the text. Right: density difference; green -- soliton region, orange -- wake region; the regions are shaded according to theoretical predictions, as an illustration of the part that belongs to the soliton and the wake.
The total $N_s+N_w=3.000$ is conserved.}
\label{fig3}
\end{center}
\end{figure}

For numerical calculations, we slightly smoothed the edges (see Figure \ref{fig1}) so that the modulation instability develops more slowly and the soliton has enough time to move along the smooth background. The dynamics of the soliton can be found from the equations \eqref{eq7} and \eqref{eq8}, where the previously found solution for the background wave should be substituted. The background parameters are $a=1,l=20$ and the soliton is set at the initial coordinate $x_0=-28$ near the hump. The comparison of the numerical calculation and the theoretical trajectory for different speeds and half-widths of the soliton is presented in Figure \ref{fig2}, where it can be seen that even for ``wide'' solitons the agreement is very good. In the numerical calculation we made two runs with the same initial hump, with and without the soliton. The densities of the two runs were subtracted, and the position of the soliton was taken as the centre of $(|\psi|^2-|\psi_b|^2)^2$ in a window around the soliton (see the right column of Figure~\ref{fig3}), so the small oscillations behind it are not counted.

It is interesting to see what happens to the norm and phase accumulation during the passage of the soliton. In the second column of Figure~\ref{fig3} the density difference $|\psi|^2-|\psi_b|^2$ is shown at different moments in time. Having calculated the theoretical values \(N_s\) and \(N\) using the formulas from the previous section, we shaded in orange the part of the hump that the soliton leaves behind, \(N-N_s\), and in green the soliton's own norm value \(N_s\). The boundary between the soliton and its wake cannot be defined uniquely, since the wake is attached to the soliton without a gap; we put it at the last point behind the soliton where the norm of the wake counted from the left is equal to $N-N_s$.

An interesting picture is observed in the left column of Figure \ref{fig3}. To the left of the soliton, there is a significant difference between the background with the soliton $\psi$ and the background without the soliton $\psi_b$. This view clearly shows that the background points through which the soliton has already passed have acquired a phase shift. The options for its theoretical calculation are shown with two dashed lines. The black line shows 
\begin{equation}\label{ex4}
    |\psi-\psi_b|=|\psi_b||e^{i\Delta\vartheta}-1|=2\sqrt{\rho_b}|\sin{2\alpha}|
\end{equation}
where $\Delta\vartheta=4\alpha$ and $\rho_b(x,t)$ are taken according to the theoretical formulas at the moment in time indicated in the picture ($t=5$ and $t=9$). 
The orange dotted line marks a slightly different approach, based on the idea that the background points through which the soliton passes carry with them the shifted phase, like Lagrangian particles. According to hydrodynamics, each particle moves with the flow velocity $u$. From this and the continuity equation \eqref{eq1}, it follows that a natural Lagrangian variable (that is, the one that is transported with the particle) is the mass of the fluid to the left of the particle $m = \int_{-\infty}^x\rho dx'$, since $m_t+um_x=0$. From the flow formula \eqref{ex2}, it also follows for particles that
\begin{equation}\label{ex5}
    u=\frac{dX}{dt} = \frac{f'}{f}X \quad\rightarrow\quad X(t)=f(t)x_0
\end{equation}
meaning that another dimensionless invariant carried by the particle is $\xi=x/lf(t)$. From Talanov's solution \eqref{ex2}, it also follows that the invariant for the particle is $\rho f = a^2(1-\xi^2)$. Let the soliton approach the hump at \(t_0\), and we draw the graph for time \(t\) with the encounter with each particle on the way occurring at the moment $\tau$ (\(t_0<\tau<t\)) and coordinate $x_s(\tau)$. Then, from the invariants \( x/lf \) and \( f\rho \), it follows that the graph according to formula \eqref{ex4} for the Lagrangian representation can be easily found in parametric form
\begin{equation}\label{ex6}
    x(\tau)=x_s(\tau)\frac{f(t)}{f(\tau)}, \qquad \rho(\tau)=\rho_b\big(x_s(\tau),\tau\big)\frac{f(\tau)}{f(t)},
\end{equation}
where $t_0<\tau<t$, $t$ is the moment shown in the figure, $x_s(\tau)$ is the known trajectory of the soliton, $\rho_b(x,t)$ is the Talanov solution \eqref{ex2} and $\alpha = \alpha(\tau)$; the curve is plotted parametrically as $\big(x(\tau),\,2\sqrt{\rho(\tau)}\,|\sin2\alpha(\tau)|\big)$. As can be seen from Figure~\ref{fig3}, the Lagrangian representation agrees with the numerical experiment better than the black curve with $\alpha$ taken at the current moment. In the same figure, the left tail that does not align with the theoretical curve is caused by the waves going from the edges of the hump inside it.

\section{Conclusion}

We obtained a convenient formula for describing the trajectory of solitons \eqref{eq9} along a large-scale background wave and the corresponding Hamiltonian description for the NLS equation with focusing. We generalized Stokes' observation of the connection between the soliton tail and linear waves to the case of a breather. Thus, following Sazonov's works on complex wave numbers and the frequency of bright solitons, we obtained a convenient formula \eqref{eq8} for describing the trajectory of the soliton, knowing the dispersionless evolution of the background parameters. As a result, it was shown that the breather has a complex parameter \( q \) that is conserved along its motion. Based on this, we constructed a Hamiltonian theory for the soliton \eqref{ham8}, which required the introduction of a complex angle $\Phi = \alpha+i\beta$ to obtain explicit dependencies \eqref{ham10}. They agree with the renormalized integrals of the Tajiri--Watanabe breather, and the angle $\alpha$ gets a simple meaning, $4\alpha$ is the jump of the background phase across the soliton.

When the soliton passes a point of the background, the background behind it is rotated by $\Delta\vartheta=4\alpha$, where $\alpha$ is taken at the moment of passage. On a ring this phase jump has to be compensated by a counterflow, but on a hump that goes to zero at its edges there is no counterflow, and the background behind the soliton simply remains rotated. Since $\alpha$ changes along the path, different parts of the background get different phase shifts. After that the shifted phase does not stay at the same point of space and moves with the flow. We showed how one can approach the description of the evolution $\Delta \vartheta$ through the Lagrangian description of fluid particles.

Also, from Hamiltonian mechanics, the conservation of action $N=4\im q$ was demonstrated and it is not the norm of the soliton. When the soliton goes into a region with nonzero background, its norm decreases to $2\im K$, the rest \eqref{ham14} is left in the wake, and the soliton takes it back when it returns to a zero background. A soliton-generated wake was found also in~\cite{BiondiniLiMantzavinos2018} for the structures produced by the modulation instability, but there the amount of norm in it was not determined.

The proposed theory works only in regions of background large-scale waves where there is no developed modulational instability yet.

\section*{Declaration of competing interest}
The authors declare that they have no known competing financial interests or personal relationships that could have appeared to influence
the work reported in this paper.

\section*{Acknowledgements}
I am very grateful to A.V. Slyunyaev and S.V. Sazonov for the problem they posed. I also want to thank S.V. Sazonov for introducing me to his articles, which made this work possible.

\section*{Data availability}
Data will be made available on request.

\section*{Declaration of generative AI and AI-assisted technologies in the writing process}
During the preparation of this work, the author used Claude (Anthropic) for editing the language of the manuscript, assisting with the literature for the introduction and writing scripts for numerical simulations and figures.

\appendix

\section{Breather on a uniform background}\label{app:breather}

The breather on a uniform background was found by Tajiri and Watanabe~\cite{TajiriWatanabe1998} for the equation
$iu_t+u_{xx}+\gamma|u|^2u=0$ in the form $u=\rho_0e^{i\theta}g/f$,
Eqs.~(9)--(11) of~\cite{TajiriWatanabe1998}. Equation~(1) corresponds to $\gamma=2$ with the time rescaled by a factor of two, $\psi(x,t)=u(x,t/2)$; their background amplitude and wave number are
$\rho_0=\sqrt\rho$ and $k=u$, and their complex angle is $\phi_1=-\Phi^*=-\alpha+i\beta$, where $\Phi=\alpha+i\beta$ is
the angle of Sec.~\ref{sec:ham}. In the notation of the paper the solution reads
\begin{equation}\label{B:TW}
\psi=\sqrt\rho\,e^{i\vartheta_b}\,
\frac{1+e^{-2i\Phi}E+e^{-2i\Phi^*}E^*+a\,e^{-4i\alpha}EE^*}{1+E+E^*+a\,EE^*},\qquad
a=\frac{\cosh^2\beta}{\cos^2\alpha},
\end{equation}
where $\vartheta_b=ux-(u^2/2-\rho)t$ is the phase of the background and
\begin{equation}
E=\exp\big[i(Kx-\Omega t)\big],\qquad K=2\sqrt\rho\sin\Phi,\qquad \Omega=K\Big(\frac u2+q\Big),
\end{equation}
is the tail of the soliton and $q=u/2+i\sqrt\rho\cos\Phi$. The envelope depends on $|E|=e^{-\operatorname{Im}K\,(x-Vt)}$ and moves with the velocity
\eqref{eq6} and the carrier breathes with the frequency \eqref{eq9}.

\section{Hamilton’s theory of the dark soliton}\label{A:dark}
For the defocusing case, the smooth background is described by two real Riemann invariants $r_\pm=\frac{u}2\pm\sqrt{\rho}$:
\begin{equation}\label{dark1}
\frac{\partial r_{\pm}}{\partial t}+v_\pm(r_-,r_+)\frac{\partial r_\pm}{\partial x} = 0,
\end{equation}

where $v_\pm=\frac12(3r_\pm+r_\mp)$. A dark soliton has tails $\propto e^{-\kappa|x|}$,
i.e., $K=\pm i\kappa$ and the Stokes observations follow from the integral along the path with $K^2=-\kappa^2$:
\begin{equation}\label{dark2}
\kappa^2=4(r_+-q)(q-r_-),\qquad V=\tfrac12(r_++r_-)+q,\qquad r_-<q<r_+ .
\end{equation}

Although in the canonical case it is considered that $H = H(p,x)$, in this case it will be convenient to use the canonical equations of motion 
\begin{equation}\label{dark3}
\dot{x} = V=\frac{\partial H}{\partial p}\Bigg|_x, \quad \dot p = -\frac{\partial H}{\partial x}\Bigg|_p,
\end{equation}
but to consider that the Hamiltonian implicitly depends on momentum and coordinates, namely $H = H(q,r_+,r_-)$, $p = p(q,r_-,r_+), r_\pm = r_\pm(x,t), \dot q=0$ along the path. Thus, the first Hamiltonian equation can be easily rewritten as
\begin{equation}\label{dark4}
H_q=V\,p_q.
\end{equation}
Next, we note that $dp = p_{r_-} dr_-+p_{r_+}dr_++p_qdq$ and $\dot r_\pm=(V-v_\pm)\partial_xr_\pm$, from which it follows that $\dot p=\sum_\pm p_{r_\pm}(V-v_\pm)\partial_xr_\pm$.  On the other hand,
\begin{equation}\label{dark5}
\frac{\partial H}{\partial x}\Bigg|_p = \sum_\pm\ H_{r_\pm}\partial_x r_\pm+H_q\frac{\partial q}{\partial x}\Bigg|_p
\end{equation}
where the expression for $\partial q/\partial x|_p=-\sum_\pm p_{r_\pm}\partial_xr_\pm/p_q$ can be found from the same $dp$ written above. Thus, the second Hamiltonian equation takes the form
\begin{equation}\label{dark6}
H_{r_\pm}=v_\pm\,p_{r_\pm}.
\end{equation}

Differentiating (\ref{dark4}) with respect to $r_\pm$ and (\ref{dark6}) with respect to $q$ gives $(v_\pm-V)p_{qr_\pm}=V_{r_\pm}p_q$. With $V_{r_\pm}=\frac12$ and
$v_\pm-V=r_\pm-q$,
\[
\frac{\partial}{\partial r_+}\ln p_q=\frac1{2(r_+-q)},\qquad \frac{\partial}{\partial r_-}\ln p_q=\frac1{2(r_--q)},
\]
hence $p_q=F(q)\sqrt{(r_+-q)(q-r_-)}=\frac12F(q)\kappa$.
As shown in \cite{Kamchatnov2024TMF}, to ensure that the result for homogeneous backgrounds aligns with the classical one, $F$ should be set to $-4$, so $p_q=-2\kappa$. We arrive at the conclusion that the canonical momentum and Hamiltonian of the dark soliton can be expressed in the form of integrals (with $p=0$ for a vanishing soliton $q=r_+$)
\begin{equation}\label{dark7}
p=2\int_q^{r_+}\kappa(q')\,\dd q',\qquad H=2\int_q^{r_+}V(q')\,\kappa(q')\,\dd q' .
\end{equation}

The product $(r_+-q)(q-r_-)$ is of the type $1-x^2$, and the standard way to remove $\sqrt{1-x^2}$ is $x=\cos\varphi$. Shifting $[-1,1]$ to $[r_-,r_+]$ gives
\begin{equation}\label{dark8}
\begin{split}
q=\tfrac12(r_++r_-)+\tfrac12(r_+-r_-)\cos\varphi,\qquad \kappa=(r_+-r_-)\sin\varphi,\\ V=(r_++r_-)+\tfrac12(r_+-r_-)\cos\varphi.
\end{split}
\end{equation}
The edges $\varphi=0,\pi$ ($q=r_\pm$) correspond to vanishing solitons, and $\varphi=\pi/2$ to the black soliton. The substitution (\ref{dark8}) gives $\kappa\,\dd q'=-\frac12(r_+-r_-)^2\sin^2\varphi'\dd\varphi'$ and reduces (\ref{dark7}) to $\int_0^\varphi\sin^2$ and $\int_0^\varphi\sin^2\cos$, so we obtain explicit formulas
\begin{equation}\label{dark9}
p=\tfrac12(r_+-r_-)^2(\varphi-\sin\varphi\cos\varphi),\qquad H=(r_++r_-)\,p+\tfrac16(r_+-r_-)^3\sin^3\varphi .
\end{equation}

\bibliographystyle{elsarticle-num} 
\bibliography{refs}

\end{document}